\documentclass[prl,twocolumn,amsmath,amssymb,superscriptaddress]{revtex4-1}
\usepackage[utf8]{inputenc}
\usepackage[T1]{fontenc}
\usepackage{xcolor}
\usepackage{graphicx}
\usepackage{wasysym}
\usepackage{lmodern}

\usepackage{lipsum}  

\definecolor{darkblue}{rgb}{0,0,0.6}
\definecolor{darkred}{rgb}{0.6,0,0}
\usepackage[colorlinks=true,urlcolor=darkblue,citecolor=darkblue,linkcolor=darkred]{hyperref}

\newcommand{\dd}{\mathrm{d}}

\DeclareMathOperator{\erfc}{erfc}

\def \equi#1{\mathrel{\mathop{\kern 0pt\sim}\limits_{#1}}}

\DeclareMathOperator{\sign}{sign}

\begin{document}

\title{Generalised density profiles in single-file systems}

\author{Alexis Poncet}
\affiliation{Sorbonne Universit\'e, CNRS, Laboratoire de Physique Th\'eorique de la Mati\`ere Condens\'ee (LPTMC), 4 Place Jussieu, 75005 Paris, France}

\author{Aur\'elien Grabsch}
\affiliation{Sorbonne Universit\'e, CNRS, Laboratoire de Physique Th\'eorique de la Mati\`ere Condens\'ee (LPTMC), 4 Place Jussieu, 75005 Paris, France}

\author{Pierre Illien}
\affiliation{Sorbonne Universit\'e, CNRS, Laboratoire de Physico-Chimie des \'Electrolytes et Nanosyst\`emes Interfaciaux (PHENIX), 4 Place Jussieu, 75005 Paris, France}

\author{Olivier Bénichou}
\affiliation{Sorbonne Universit\'e, CNRS, Laboratoire de Physique Th\'eorique de la Mati\`ere Condens\'ee (LPTMC), 4 Place Jussieu, 75005 Paris, France}

\date{\today}

\begin{abstract}
 Single-file diffusion refers to the motion in narrow channels of particles which cannot bypass each other, and leads to tracer subdiffusion.
 Most approaches to this  celebrated many-body problem were restricted to the description of the tracer only.
Here, we go beyond this standard description by introducing  and {providing analytical results} for generalised density profiles (GDPs) in the frame of the tracer. In addition to controlling  the statistical properties of the tracer, these quantities fully characterise the correlations between the tracer position and the bath particles density. Considering the hydrodynamic limit of the problem, we {determine} the scaling form of the GDPs with space and time, and unveil a non-monotonic dependence with the distance to the tracer despite the absence of any asymmetry. 
 Our analytical approach provides {several} exact results for the GDPs
 for paradigmatic models of single-file diffusion, such as Brownian particles with hardcore repulsion, the Symmetric Exclusion Process  and the Random Average Process.
 {The range of applicability  of our approach is further illustrated by considering (i) extensions to general interactions between particles, (ii) the out-of-equilibrium situation of an initial step of density and (iii) beyond the hydrodynamic limit, the GDPs at arbitrary time in the dense limit.}
\end{abstract}

\maketitle


The key feature of single-file diffusion \cite{Chou:2011, MALLICK201517, P.L.-Krapivski:2009}, which refers to the motion of particles which cannot bypass each other, is that a typical displacement of a tracer particle scales as $t^{1/4}$ instead of 
$t^{1/2}$ as in regular diffusion \cite{Harris:1965, Levitt:1973, Alexander:1978, Arratia:1983, Rodenbeck:1998,Taloni:2008,Taloni2011,Taloni2014}. This subdiffusive scaling has
been demonstrated in a growing number of experimental realizations, in contexts as varied as transport in porous media \cite{Gupta:1995,Meersmann:2000}, zeolites \cite{Hahn:1996} or confined colloidal suspensions \cite{Wei:2000,Lin:2005}. Theoretically, it has lead to a huge number of works in the physical and mathematical literature. Recent advances include the determination of the large deviation functions of the position of a tracer in a system of Brownian particles with hardcore repulsion  \cite{Krapivsky:2014,PhysRevLett.113.120601,Sadhu:2015} and in the Symmetric Exclusion Process \cite{Illien:2013, Imamura:2017} (see below for definitions), which are two paradigmatic models of single-file diffusion with hard-core interactions. 

Theoretical interest in single-file diffusion originates from the  challenging many-body nature of the problem: any large  displacement of the tracer particle in one direction  requires large displacements of more and more bath particles in the same direction. 
{However, the quantification of the coupling between the tracer position and the bath particles density remains a broadly open question.}
Here, we develop a theoretical framework with which to determine these correlations for  single-file systems.


\begin{figure}
\centering
\includegraphics{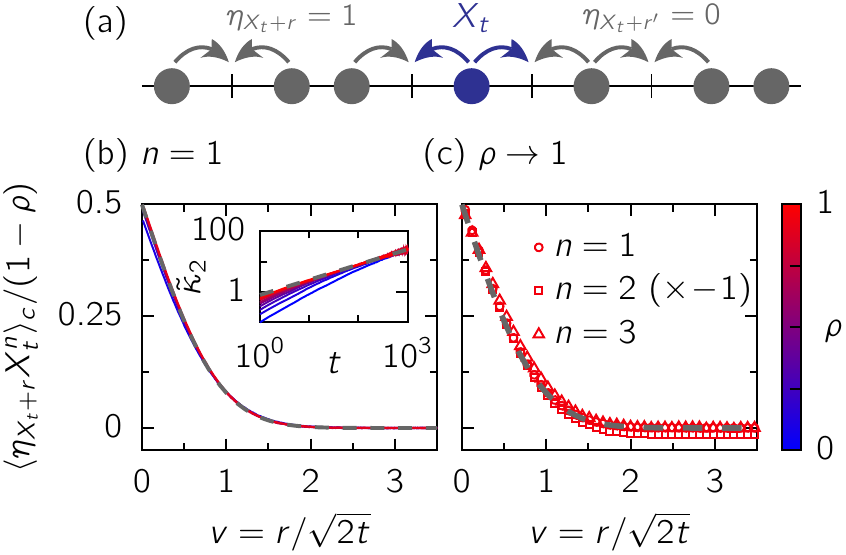}
\caption {SEP. 
{(a)  The Symmetric Exclusion Process (SEP). The position of the tracer is called 
$X_t$ and the occupation numbers of the sites with respect to the tracer are denoted $\eta_{X_t+r}$.}
(b) Profiles of order $1$ at densities $0.1, 0.25, 0.5, 0.75, 0.9$
at time $t=3000$ (blue to red).
Dashed gray line: prediction from Eq.~\eqref{eq:resPhi1}. Inset: rescaled variance $\tilde\kappa_2 = \kappa_2 \rho/(1-\rho)$ compared 
to known expression (gray) \cite{Arratia:1983} and retrieved by our approach.
(c) Profiles of order $1$, $2$ and $3$ at high density ($\rho = 0.95, t=1000)$. Dashed gray line: prediction 
from Eq.~\eqref{eq:pred_high}. 
}
\label{fig:SEP}
\end{figure}

We introduce our formalism by considering first the  Symmetric Exclusion Process (SEP). Particles, present at a density
$\rho$, perform symmetric continuous-time random walks on a
one-dimensional lattice with unit jump rate, and hard-core
exclusion is enforced by allowing at most one particle per site (Fig.~\ref{fig:SEP}(a)). The tracer,  of position $X_t$  at time $t$, is initially at the origin. The bath particles are described by the set of occupation numbers  $\eta_r(t)$ of each site $r\in\mathbb{Z}$ of the line at time $t$, with $\eta_r(t) = 1$ if the site is occupied and $\eta_r(t) = 0$ otherwise. 

{To quantify the coupling between the position of the tracer and the density of bath particles,
 we study the joint process $(X_t, \eta_{X_t+r})$, which is entirely characterized by its joint cumulant generating function $\ln \langle e^{\lambda X_t + \chi \eta_{X_t+r}} \rangle$. Since $\eta_{X_t+r}$ only takes values $0$ and $1$, this function takes the form
  \begin{equation}
    \ln \langle e^{\lambda X_t + \chi \eta_{X_t+r}} \rangle
    =
    \psi(\lambda,t) + 
    \ln (1 + (e^\chi - 1) w_r(\lambda,t)),
  \end{equation}
  where $\psi(\lambda,t)$ is the usual  cumulant-generating function, whose expansion defines the cumulants $\kappa_n$ of the position: 
\begin{equation}
\label{defpsi}
\psi(\lambda, t) \equiv \ln \left\langle e^{\lambda X_t} \right\rangle \equiv 
\sum_{n=1}^\infty \frac{\lambda^n}{n!} \kappa_n(t).
\end{equation}
In turn,  $ w_r(\lambda, t)$ is 
 the GDP-generating function \cite{Illien2015}}~\footnote{Note that the GDP-generating function~\eqref{eq:def_gdp} can be thought as the average occupation in a biaised ensemble. This type of objects has received a lot of attention over the last years~\cite{Touchette2015,Chetrite2015,Derrida2019}}.{
\begin{equation}
  \label{eq:def_gdp}
  w_r(\lambda, t) = \langle \eta_{X_t+r} e^{\lambda X_t}\rangle /\langle e^{\lambda X_t}\rangle,
\end{equation}
whose  expansion 
gives the joint 
cumulants $\langle \eta_{X_t+r} X_t^n\rangle_c$ of the tracer position $X_t$ and the occupation number $\eta_{X_t+r}$ measured in its frame of reference.
  For instance, the first cumulant $\langle \eta_{X_t+r} X_t \rangle_c = \mathrm{Cov}(\eta_{X_t+r},X_t)$ provides a measure of the correlations between the displacement of the tracer and the occupation of the site at a distance $r$ of the tracer, while the second one $\langle \eta_{X_t+r} X_t^2 \rangle_c = \mathrm{Cov}(\eta_{X_t+r},X_t^2)$ (since $\langle X_t \rangle$=0) gives a measure of the correlation between the amplitude of the fluctuations of $X_t$ and the occupation at a distance $r$ of $X_t$.
  Finally, the joint cumulant generating function $\ln \langle e^{\lambda X_t + \chi \eta_{X_t+r}} \rangle$ is given by the knowledge of $\psi$ and $w_r$, which are thus key quantities, whose joint determination is the object of this Letter.}

While it is known that the cumulant-generating function $\psi$ scales as $\sqrt{t}$  at large time for single-file systems, we put forward that, more generally,   at large scale (large time, large distances), the GDP-generating function admits the  scaling form:  
\begin{equation}
\label{scaling}
w_r(\lambda, t) - \rho \equi{t\to\infty} \Phi\left(\lambda,v=\frac{r}{\sqrt{2t}}\right)\equiv\sum_{n=1}^\infty \frac{\lambda^n}{n!} \Phi_n(v).
\end{equation}
We note that the symmetry of the system imposes $\Phi(\lambda, v) = \Phi(-\lambda, -v)$: in the following the results will be stated only for $v>0$.
When the tracer moves, it perturbs the bath particles and gives birth to  density profiles which, as displayed by this scaling form, are not stationary but involve typical length scales growing as $\sqrt{t}$. In turn, the way the bath particles readjust at the front and behind the tracer (encoded in $w_{\pm 1}$) controls its displacement, leading to (see Supplementary Material, SM \cite{SM}) 
\begin{equation} \label{eq:link_psi_w}
\frac{\mathrm{d}\psi}{\mathrm{d} t} = \frac{1}{2} \left[ (e^\lambda - 1)(1-w_1) + (e^{-\lambda} - 1)(1-w_{-1}) \right].
\end{equation}
Finally, besides fully quantifying the  correlations between the tracer position and the density of bath particles, 
the GDPs control the time evolution of the cumulant-generating function.
{
  In particular,  the scaling $\psi(\lambda,t) \underset{t \to \infty}{\sim} \sqrt{t}$ of the cumulants \cite{Illien:2013,Imamura:2017} actually originates from the scaling form~\eqref{scaling} of the GDP generating-function.}
Relying on a mixed Eulerian (for the bath particles) and Lagrangian (for the tracer)  master equation and on the scaling of $\Phi$ (Eq. \eqref{scaling}), 
we show in SM that this key observable satisfies the hydrodynamic equation 
\begin{eqnarray}
\label{magicsystem}
\Phi''(v) + 2(v+b_\mu)\Phi'(v) + C(v) &=& 0
\end{eqnarray}
completed by the boundary conditions
\begin{align}
&\Phi'(0^\pm)+2b_\pm[\rho+\Phi(0^\pm)]  =  0 \label{eq_boundderiv}\\
&1-\rho - \Phi(0^-) = e^\lambda (1-\rho - \Phi(0^+)) \label{eq_dpsidt}
\end{align}
in front and past the tracer, 
with $\mu$ the sign of $v$, $b_\mu(\lambda)\equiv \lim_{t\to\infty}\psi(\lambda,t)/[\sqrt{2t}(e^{\mu \lambda}-1)]$ and the dependence on $\lambda$ omitted for simplicity. Equation \eqref{eq_dpsidt} comes from the cancellation of $\frac{\dd \psi}{\dd t}$ at large times (see Eq. \eqref{eq:link_psi_w}).   The function  $C(v)$ involves higher order correlations, and is {\it a priori} unknown. However, as we report below (see SM for details), explicit exact expressions of the GDPs can be obtained {in several important situations}. 

\begin{figure*}
\centering
\includegraphics{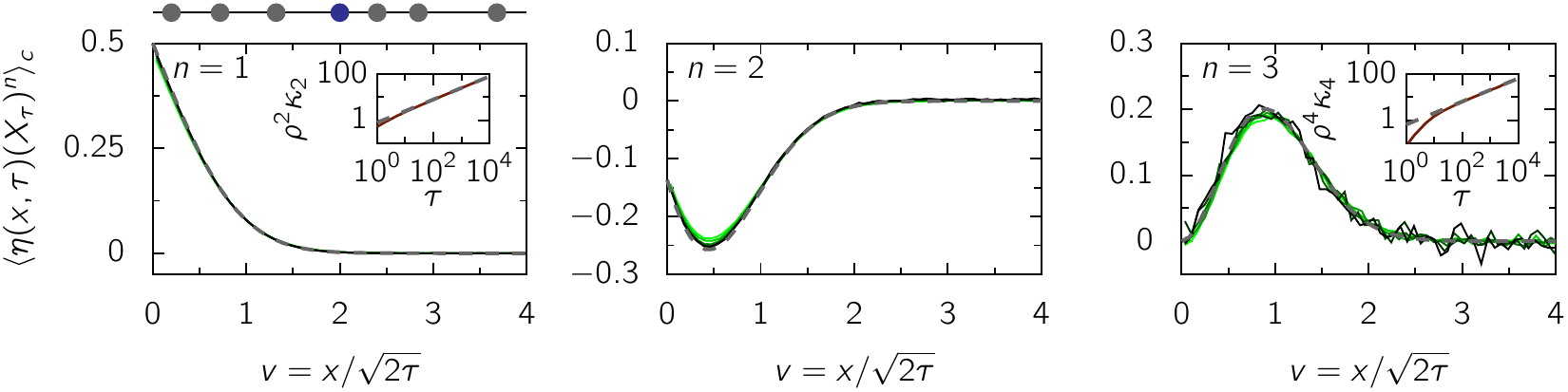}
\caption{Pointlike interacting particles. Rescaled orders 1, 2 and 3 (left to right) of the GDPs
at times $\tau = 100, 200, 500, 1000, 2000, 5000$ (green to black). The dashed gray lines are the predictions from Eqs.~\eqref{eq:resPhi1}, \eqref{eq:resPhi2Low}, \eqref{eq:resPhi3Low}.
The insets show the second and fourth cumulants with the solid lines corresponding to the simulations and the dashed gray line to the solution from Eq.~\eqref{eq:implicite}.
}
\label{fig:points}
\end{figure*}

First, we show that the function $C(v)$ is strictly equal to zero at first order in $\lambda$, making the equation \eqref{magicsystem} closed at this order, and leading to
\begin{equation}
\label{eq:resPhi1}
\Phi_1(v > 0) = \frac{1-\rho}{2} \erfc(v).
\end{equation}
This expression provides the exact large-time behavior of the correlation function  {$\langle \eta_{X_{t}+r}X_{t} \rangle_c = \mathrm{Cov}(\eta_{X_t+r},X_t)$ of the SEP at any density. The fact that it is positive for $v>0$ indicates that, if $X_t>0$, the sites to the right of the tracer have higher occupation numbers, which shows that there is an accumulation of particles in front of the tracer. } Note that it decays monotonically to zero with the distance to the tracer (Fig.~\ref{fig:SEP}(b)).  {We finally stress that \eqref{eq:resPhi1} together with \eqref{eq_dpsidt} allows one to recover in a straightforward way the well known expression of the second cumulant $\kappa_2(t)=(1-\rho)/\rho\sqrt{2t/\pi}$ \cite{Harris:1965}.}

Second, in the dense limit $\rho\to1$, it is shown that the function $C(v)$, which involves a product of occupation numbers of bath particles, vanishes,  which leads to the full GDP-generating function
\begin{equation}
\label{eq:pred_high}
\lim_{\rho\to 1}\frac{\Phi(\lambda,v>0)}{1-\rho} = \frac{1-e^{-\lambda}}{2} \erfc(v),
\end{equation}
where the dependencies on $\lambda$ and $v$ factorise. 
This expression gives the GDPs of arbitrary order, which, up to a sign,  assume all the same value, and again decay monotonically to zero (Fig.~\ref{fig:SEP}(c)). 

\begin{figure*}
\centering
\includegraphics{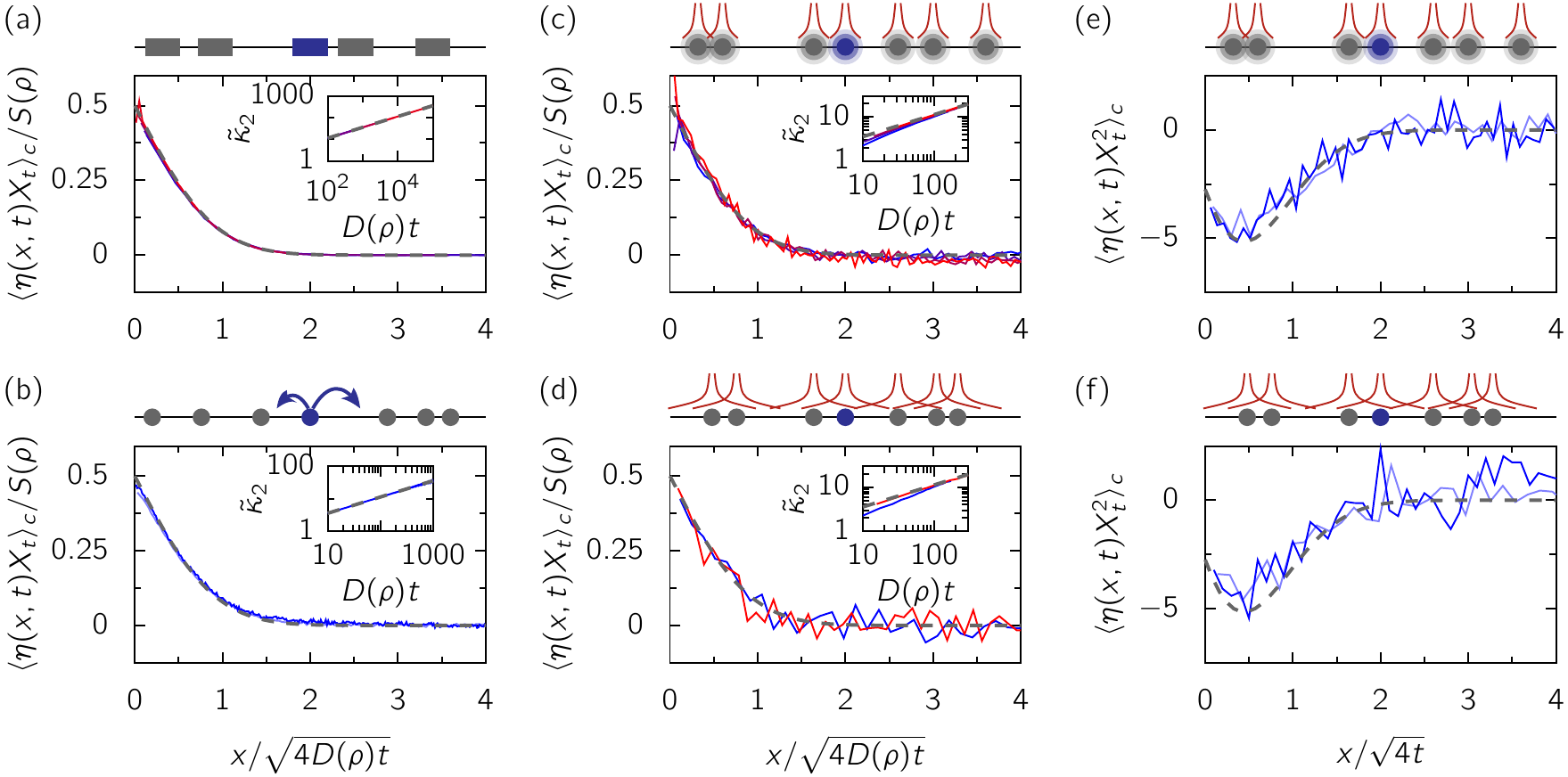}
\caption {GDPs of single-file systems.
(a-d) Profiles at order 1 for various models. The dashed gray lines correspond to the prediction of Eq.~\eqref{eq:phi1_sf}. The insets show the rescaled variance $\tilde\kappa_2 = \kappa_2 \rho/S(\rho)$ with the prediction in gray. (a) Hard-rod gas at density $\rho = 1$ and time $t=1000$. The length of a rod is $a = 0.1, 0.25, 0.5, 0.75, 0.9$ (blue to red). The parameters are $S(\rho) = (1-a\rho)^2$, $D(\rho) = (1-a\rho)^{-2}$~\cite{Lin:2005}. (b) Random-average process~\cite{Rajesh:2001, Kundu:2016} at density $\rho=1$ at times $t=1000$ and $5000$ (light blue and blue): at exponential times, each particle performs a symmetric jump whose length is a random fraction of the distance to the nearest particle. $S(\rho)$ and $D(\rho)$ are given in Ref.~\cite{Kundu:2016}. (c)  Point-like Brownian particles interacting by a Weeks-Chandler-Andersen potential {($V(r)\propto (\frac{1}{r^{12}}-\frac{1}{r^6})$ for $r<2^{1/6}$ and $0$ otherwise)}. Density $\rho = 0.2, 0.3, 0.4, 0.5$ (blue to red) at time $t=100$.
$S(\rho)$ is the structure factor at vanishing wave-number, and $D(\rho) = D_0/S(\rho)$~\cite{Kollmann2003,Lin:2005}. (d) Point-like Brownian particles interacting with long-range dipole-dipole interactions {$V(r)\propto \frac{1}{r^3}$} at density $\rho = 0.2$ and $0.4$ (blue and red) at time $t=100$. (e-f) Profiles of order 2 for the same models as (c-d) at density $\rho = 0.05$ and times $5\cdot 10^3$ and $1\cdot 10^4$ (light blue and blue).
The dashed gray line is the low-density prediction from Eq.~\eqref{eq:resPhi2Low}.
}
\label{fig:SF}
\end{figure*}
Third we investigate the dilute limit $\rho\to 0$ of the SEP.
{
  We stress that the results in this case cannot be deduced from the dense limit discussed above, since the particle-hole symmetry of the SEP is explicitly broken by choosing to follow the motion of one particle. Actually, this limits exhibits a richer phenomenology.
  In fact, it corresponds to the model of interacting point-like particles on a line~\cite{Sadhu:2015, Krapivsky:2014, PhysRevLett.113.120601}, and needs to be taken at constant $x=\rho r$ and $\tau=\rho^2 t$.}
The density field $\eta_{X_t + r}(t) \mapsto \eta(x, \tau)$ becomes continuous in space and the diffusive scaling for the GDPs reads $v = r/\sqrt{2t} = x/\sqrt{2\tau}$,
{
  leading to the definition $\hat\Phi(\hat\lambda, v) = \lim_{\rho\to 0} [\Phi(\lambda=\rho\hat\lambda, v)/\rho]$.
}
In this case, the function $C(v)$ is not negligible, but we put forward
{
  self-consistently, in order to retrieve the known cumulants $\kappa_n$ (see point (i) below), the closure relation
  \begin{equation}
    \label{closure_relation}
    \lim_{\rho \to 0} [C(\lambda = \rho \hat{\lambda},v)/\rho]
    = 2 \hat{\lambda} \frac{d \beta}{d \hat{\lambda}} \hat{\Phi}'(v),
  \end{equation}
  where we defined $\beta = \lim_{t\to\infty}\frac{1}{\sqrt{2t}}\frac{\hat\psi}{\hat\lambda}$, with $\hat\psi(\hat\lambda)=\lim_{\rho\to 0} [\psi(\lambda=\rho\hat\lambda)/\rho]$. This leads to the key and strikingly simple  closed equation for the full GDP-generating function
  \begin{equation}
    \hat\Phi''(v) + 2(v+\xi) \hat\Phi'(v) = 0,
    \label{phi_dilue_eq}
  \end{equation}
  which is the main result of this Letter. It yields
  \begin{equation}
    \hat \Phi(\hat\lambda,v > 0) = \frac{\beta\erfc(v+\xi)}{\pi^{-1/2}e^{-\xi^2}- \beta\erfc(\xi)}.
    \label{phidilue}
  \end{equation}
  where $\xi = \lim_{t\to\infty}\frac{1}{\sqrt{2t}}\frac{d\hat\psi}{d\hat\lambda}$.
  The quantities $\beta$ and $\xi = \frac{d}{d\hat{\lambda}}(\hat{\lambda} \beta)$ are then determined from Eq. \eqref{eq_dpsidt}, which becomes}
\begin{equation}
\label{eq:implicite}
\beta \sum_{\mu=\pm 1} \frac{\erfc(\mu\xi)}{\pi^{-1/2}e^{-\xi^2}-\mu\beta\erfc(\mu\xi)}
= \hat \lambda.
\end{equation}
Several comments are in order. (i) Eq.~\eqref{eq:implicite} is an implicit equation that allows one to retrieve  the exact cumulants $\kappa_n$ obtained in previous studies~\cite{Sadhu:2015, Krapivsky:2014, PhysRevLett.113.120601} (see SM \cite{SM} for detailed correspondence).
{This validates self-consistently the closure relation~\eqref{closure_relation}.}
(ii) 
Then, Eq.~\eqref{phidilue}  provides the
GDPs of interacting point-like particles at arbitrary order;
for instance (Fig.~\ref{fig:points}),
\begin{align}
\Phi_{2}(v) &=\frac{1}{\rho}\left[ \frac{1}{2}\erfc v - 2\frac{e^{-v^2}}{\pi}\right],  \label{eq:resPhi2Low} \\
\label{eq:resPhi3Low}
\Phi_{3}(v) &= \frac{3}{\pi^{3/2}\rho^2}\left[(2v - \sqrt\pi)e^{-v^2} + \sqrt\pi \erfc v \right].
\end{align}
(iii) {The sign of these GDPs is non trivial. For instance,  $\Phi_2(v) = \mathrm{Cov}(\eta_{X_t+r},X_t^2)$ is negative (see Fig.~\ref{fig:points} for $n=2$), which implies that $X_t^2$ and $\eta_{X_t+r}$ fluctuate in opposite directions: a larger fluctuation of $X_t^2$ is associated with a smaller value of the occupation. Furthermore, even if there is no asymmetry in the dynamics, these GDPs display a surprising non-monotonic behavior with the distance to the tracer (see Fig.~\ref{fig:points} for $n\geq 2$), which indicates that this effect is more pronounced at a certain distance of the tracer (which is non stationary, and grows as $\sqrt{t}$)}. {(iv) It is demonstrated in SM \cite{SM} that the GDPs  actually coincide with the saddle-point solution in the formalism of Macroscopic Fluctuation Theory \cite{Krapivsky2015}, evaluated at a specific point. In turn, it sheds new light on this saddle-point solution. {This, after (i), is a further validation of the exactness of the closure relation~\eqref{closure_relation}.} A key result of our approach is that the GDPs satisfy the very simple closed equation~\eqref{phi_dilue_eq}. (v) We stress that our approach   can be  generalized to the important out-of-equilibrium situation of an initial step of density \cite{Derrida2009,Imamura:2017}.
{The main equation~\eqref{phi_dilue_eq} remains valid in this case, and only the boundary conditions~(\ref{eq_boundderiv},\ref{eq_dpsidt}) must be straightforwardly adapted. The modified equations, and explicit}
expressions of the cumulants and the GDPs are given in SM \cite{SM} {(Eqs.~(S55-S62))}.

These analytical results can also be extended to a general  single-file system of interacting particles with average density $\rho$.
Such a system can be described at large scale by two quantities: the collective diffusion coefficient $D(\rho)$ and the static structure factor at vanishing wavenumber $S(\rho)$~\cite{Kollmann2003,Lin:2005,Krapivsky:2015}.
The case of the SEP considered above corresponds to $D(\rho) = 1/2$ and $S(\rho) = 1-\rho$.
{
  We conjecture that the first-order GDP of a general single-file system can be obtained by adapting the main equation~(\ref{phi_dilue_eq}) into
$D(\rho) \Phi_1''(v) + v \Phi_1'(v) = 0$, 
  with the boundary conditions
$   2  D(\rho)\Phi_1'(0^\pm) + \rho\tilde\kappa_2 = 0$ and 
    $\Phi_1(0^+) - \Phi_1(0^-)= S(\rho)$.
  This leads to the general expression
}
\begin{equation}
\label{eq:phi1_sf}
\Phi_1(v \ge 0) = \frac{S(\rho)}{2}
\erfc\left(\frac{v}{\sqrt{2D(\rho)}}\right).
\end{equation}
Note that  the known expression of the variance of the tracer
$\kappa_2(t) = (S(\rho)/\rho) \sqrt{4D(\rho)t/\pi}$ \cite{Kollmann2003,Krapivsky:2015} is recovered from the particular value $\Phi_1(0)$. 
The analytical result {\eqref{eq:phi1_sf} can be retrieved by a MFT computation provided in SM \cite{SM}. Furthermore, it is supported} by numerical simulations (Fig.~\ref{fig:SF})
of several paradigmatic examples of  single-file systems (see  SM \cite{SM} for definitions) with hard-core interactions (the Brownian hard-rods model, Fig.~\ref{fig:SF}(a), as involved in the experimental realisation of the quasi-1D colloidal suspension from \cite{Lin:2005}, and the random average process, Fig.~\ref{fig:SF}(b)~\cite{Ferrari:1998,Schutz:2000,Rajesh:2001,Kundu:2016})  and more general pairwise interactions (Brownian point-like particles with Weeks-Chandler-Anderson (WCA)
potential Fig.~\ref{fig:SF}(c), and dipole-dipole interactions, Fig.~\ref{fig:SF}(d), as involved in the experimental realisation of paramagnetic colloids confined in a 1D channel from \cite{Wei:2000}).

Higher order GDPs  can also be obtained in the dilute limit. Indeed, in the limit $\rho\to0$  of a generic single-file system, the collective diffusion coefficient and the structure factor should respectively satisfy $D(\rho)\to D_0$ and $S(\rho)\to 1$ where $D_0$ is the diffusion constant of an individual particle and $1$ is the structure factor of the ideal gas. Equations \eqref{phidilue}-\eqref{eq:resPhi3Low} are therefore valid for such a system at low enough density,
as confirmed by numerical simulations (Fig.~\ref{fig:SF}(e)-(f)).

{

  Although we mostly focused on the hydrodynamic limit, we stress that our approach also provides a framework to analyse the GDP-generating function~\eqref{eq:def_gdp} at all times. Starting from the microscopic equation satisfied by $w_r$ (see Eq.(S66)-(S68) in SM),  the  dense limit of the GDP-generating function  $\check w_r = \lim_{\rho \to 1} (w_r - \rho)/(1-\rho)$ in the Laplace domain  reads
  \begin{equation} \label{eq:decoup_solWr_high}
    \tilde w_r(u) =
    \int_0^{\infty} e^{-u t} \check{w}_r(t) d t
    = \frac{1}{u} \frac{1 - e^{-\nu\lambda}}{1+\alpha} \alpha^{|r|},
  \end{equation}
  where $\nu = \sign{r}$ and $\alpha = 1+u-\sqrt{(1+u)^2-1}$,
and the dense limit of the cumulant generating function at all times  $\check \psi(\lambda,t) = \lim_{\rho \to 1} \psi/(1-\rho)=te^{-t}(I_0(t)+I_1(t))[\cosh \lambda -1]$, where $I_n$ is a modified Bessel function of order $n$.
}

All together, the theoretical framework developed in this Letter allows one to quantify the correlations between the tracer and the bath of particles in single-file diffusion with the help of generalized density profiles.
{We emphasize the main merits of this method. First, it is based on a master equation, which is a natural tool for physicists. Second, in the limits considered here, it reduces to a simple first order linear differential equation. In views of the complexity of available methods to study tracer diffusion in single-file systems, such as coupled nonlinear partial differential equation for the Macroscopic Fluctuation Theory~\cite{Krapivsky:2014,Krapivsky2015} or integrable probabilities and Bethe ansatz~\cite{Imamura:2017}, this is an important simplification. Furthermore, we discover a closure relation which allows to break the infinite hierarchy in a many-body model of point-like particles with hardcore repulsion. Beyond these technical aspects, the impact of the present results is further demonstrated by considering  extensions to general interactions between particles,  the out-of-equilibrium situation of an initial step of density and,  beyond the hydrodynamic limit, the GDPs at arbitrary time in the dense limit. Finally, our method opens the way to the resolution at arbitrary density, which is a long-standing and challenging question. As a first step in this direction, we emphasize that the boundary equation \eqref{eq_boundderiv} remarkably holds at {\it any} density. }


\bibliographystyle{apsrev4-1}
\bibliography{listeoverleaf}

\end{document}


\title{Generalised density profiles in single-file systems\texorpdfstring{\\}{--} Supplementary Material}

\maketitle

\renewcommand{\theequation}{S\arabic{equation}}
\renewcommand{\thefigure}{S\arabic{figure}}

\tableofcontents

\section{Equations}
\subsection{Master equation of the SEP}
We consider the symmetric exclusion process (SEP) with a
tracer. The position of the tracer is denoted $X$ and the configuration of the system is denoted $\ueta = \{\eta_r\}_{r\in\mathbb{Z}}$ where $\eta_r \in {\{0, 1\}}$ is the occupation of site $r$ ($1$ if the site is occupied, $0$ if it is empty). At time $t$, the system is characterized by a probability law $P(X, \ueta, t)$.

The initial conditions are given by the equilibrium probability law for the occupations with the tracer at position $0$,
\begin{equation}
P(X, \ueta, 0) = \delta_{X, 0} \delta_{\eta_0, 1} \prod_{r\in\mathbb{Z}^\ast} \delta_{\eta_r, \gamma_r}.
\end{equation} 
where $\gamma_r$ are independent Bernouilli variables with parameter $\rho$ (density of the system).

One checks that the time-evolution of the tracer and the bath is given by the following master equation,
\begin{align} \label{eq:sm_master}
\partial_t P(X, \ueta, t) =&~ \frac{1}{2}\sum_{r\neq X, X-1} \left[P(X, \ueta^{r,+}, t)-P(X, \ueta, t)\right] \nonumber \\
&+ \frac{1}{2} \sum_{\mu=\pm 1} \left\{
(1 - \eta_X) P(X-\mu, \ueta, t) - (1-\eta_{X+\mu}) P(X, \ueta, t) \right\}.
\end{align}

The first term corresponds to the jumps of the bath particles while the second one takes into account the jumps of the tracer. We call $\ueta^{r,+}$ the configuration $\ueta$ in which the occupations of sites $r$ and $r+1$ are exchanged.

If one considers an observable $O(X, \ueta)$, its average at time $t$ is defined as
\begin{equation} \label{eq:sm_obs}
 \langle O\rangle(t) \equiv \sum_{X, \ueta} O(X, \ueta) P(X, \ueta, t).
\end{equation}
The time evolution of this average can be computed using the master equation~\eqref{eq:sm_master}.

\subsection{Observables and large-times scalings} \label{ss:sm_obs}
The first observable that we compute is the cumulant-generating function of the displacement of the tracer,
\begin{equation} \label{eq:sm_def_psi}
\psi(\lambda, t) \equiv \ln \left\langle e^{\lambda X} \right\rangle.
\end{equation}
Its expansion in powers of $\lambda$ generates the cumulants of the tracer.
At large time $t$, it scales as $\sqrt{t}$,
\begin{equation}
\psi(\lambda, t) \equi{t\to\infty} A(\lambda)\sqrt{2t}.
\end{equation}

The second observable corresponds to the generalized profiles,
\begin{equation} \label{eq:sm_def_wr}
w_r(\lambda, t) \equiv \frac{\langle \eta_{X+r} e^{\lambda X}\rangle}
{\langle e^{\lambda X}\rangle}.
\end{equation}
The expansion in powers of $\lambda$ gives the cross-cumulants between the occupations and the displacement of the tracer. At large time, they satisfy a diffusive scaling $r/\sqrt{t}$,
\begin{equation} \label{eq:sm_scale_w}
w_r(\lambda, t) - \rho \equi{t\to\infty}  \Phi\left(v = \frac{r}{\sqrt{2t}}, \lambda\right)
\end{equation}

Finally, we consider the ``modified centered correlations'',
\begin{equation} \label{eq:sm_def_f}
f_{\mu, r}(\lambda, t) \equiv \displaystyle
\frac{\left\langle(1-\eta_{X+\mu})\eta_{X+r}e^{\lambda X}\right\rangle}{\langle e^{\lambda X}\rangle} - 
\begin{cases}
 (1-w_\mu) w_{r-\mu} & \text{ if } \mu r > 0 \\
 (1-w_\mu) w_r  & \text{ if } \mu r < 0
\end{cases}.
\end{equation}
At large time, the leading term is in $t^{-1/2}$ and the sub-leading term in $t^{-1}$ with the same diffusive scaling as for the profiles,
\begin{equation} \label{eq:sm_scale_f}
f_{\mu, r}(\lambda, t) = \frac{1}{\sqrt{2t}} F_\mu\left(v=\frac{r}{\sqrt{2t}}, \lambda\right)
+ \frac{1}{2t} G_\mu\left(v=\frac{r}{\sqrt{2t}}, \lambda\right) + \mathcal{O}(t^{-3/2}).
\end{equation}

\subsection{Equations at arbitrary time} \label{ss:sm_eqs}
Using Eqs.~\eqref{eq:sm_master} and \eqref{eq:sm_obs}, one obtains the following equations for the time-evolution of the cumulant-generating function and of the generalized profiles.
\begin{align}
\label{eq:sm_evolPsi}
\partial_t \psi &= \frac{1}{2}\left\{ (e^\lambda-1)(1-w_1) + (e^{-\lambda}-1)(1-w_{-1}) \right\}, \\
\label{eq:sm_decoup_1}
\partial_t w_r &=
\frac{1}{2} \Delta w_r
- B_\nu \nabla_{-\nu}w_r
+ \frac{1}{2} \sum_{\mu=\pm 1} \left(
e^{\mu\lambda} f_{\mu, r+\mu} - f_{\mu, r}
\right) \qquad (r\neq \pm 1) \\
\label{eq:decoup_2}
\partial_t w_{\pm 1} &= \frac{1}{2} \nabla_\pm w_{\pm 1} + B_{\pm} w_{\pm 1} + \frac{1}{2} \left( e^{\pm\lambda} f_{\pm 1, \pm 2} - f_{\mp 1, \pm 1} \right)
\end{align}
$\nu$ is the sign of $r$, the gradients are $\nabla_\mu u_r = u_{r+\mu} - u_r$, and
\begin{equation}
B_{\pm} = \frac{\partial_t \psi}{e^{\pm\lambda} - 1}.
\end{equation}
In addition to that, the generalized profiles at large distance are equal to the density 
$\lim_{r\to\pm\infty}w_r = \rho$. 

\subsection{Hydrodynamic equations at large time} \label{ss:sm_eqs2}
Using the time scalings of subsection~\ref{ss:sm_obs} into the equations of subsection~\ref{ss:sm_eqs}, we first obtain two ``symmetry'' relations
\begin{align}
\label{eq:sm_decoup_3b}
1-\rho - \Phi(0^-) &= e^\lambda (1-\rho - \Phi(0^+)),\\
F_{-1}(v) &= e^\lambda F_1(v).
\end{align}

Then, we obtain the following hydrodynamic equations for the generalized profiles,
\begin{align}
\label{eq:sm_decoup_1b}
&\Phi''(v) + 2(v+ b_\nu) \Phi'(v) + C(v) = 0, \\
&C(v) = (e^\lambda - 1)F_1'(v) + \sum_{\mu=\pm 1} (e^{\mu\lambda}-1) G_\mu(v)\\ 
\label{eq:sm_decoup_2b}
&\Phi'(0^\pm) + 2b_\pm[\rho + \Phi(0^\pm)] = 0, \\
\label{eq:sm_decoup_4b}
&\Phi(v) \xrightarrow[v\to\pm\infty]{} 0,
\end{align}
with $\nu$ the sign of $v$ and $b_\pm(\lambda) = \sqrt{2t} B_\pm(\lambda) = \pm A(\lambda)/(e^{\pm\lambda} - 1)$.

\section{Results}
\subsection{First order}
At order $1$ in $\lambda$, $\Phi(\lambda, v) = \lambda \Phi_1(v) + \mathcal{O}(\lambda^2)$ and $b_\pm = \lambda \tilde\kappa_2 / 2  + \mathcal{O}(\lambda^2)$ with $\hat\kappa_2 = [\langle X^2(t)\rangle - \langle X(t)\rangle^2] / \sqrt{2t}$. Equations \eqref{eq:sm_decoup_3b}, \eqref{eq:sm_decoup_1b}, \eqref{eq:sm_decoup_2b} and \eqref{eq:sm_decoup_4b} lead to
\begin{align}
\label{eq:sm_decoup_1d}
\Phi_1''(v) + 2v \Phi_1'(v) = 0, \\
\Phi_1'(0^\pm) + \rho\tilde\kappa_2 = 0, \\
\Phi_1(0^+) - \Phi_1(0^-)= 1-\rho, \\
\label{eq:sm_decoup_4d}
\Phi_1(v) \xrightarrow[v\pm\infty]{} 0.
\end{align}

This set of equations is closed. Its resolution gives the generalized profiles at order $1$ and the rescaled variance of the tracer,
\begin{align}
\Phi_1(v\gtrless 0)  &= \pm \frac{1-\rho}{2} \erfc |v|, \\
\tilde\kappa_2 &= \frac{1-\rho}{\rho} \frac{1}{\sqrt\pi}.
\end{align}
The result for the variance is the well-known one~\cite{Arratia_1983}.

\subsection{High density}
We now turn to the high density limit $\rho\to 1$. We define the following quantities (that no longer depend on the density),
\begin{align}
\check\Phi(v) &= \lim_{\rho\to 1} \frac{\Phi(v)}{1-\rho}, &
\check b_\pm &= \lim_{\rho\to 1} \frac{b_\pm}{1-\rho}. 
\end{align}
Equations \eqref{eq:sm_decoup_3b}, \eqref{eq:sm_decoup_1b}, \eqref{eq:sm_decoup_2b} and \eqref{eq:sm_decoup_4b} simplify into 
\begin{align}
\check \Phi''(v) + 2v \check \Phi'(v) = 0, \\
\check \Phi_1'(0^\pm) + 2\check b_\pm = 0, \\
e^\lambda[1-\check \Phi(0^+)] = 1-\check \Phi(0^-), \\
\check \Phi(v) \xrightarrow[v\pm\infty]{} 0.
\end{align}
The set of equations is closed and leads to
\begin{align}
\check \Phi(v \gtrless 0) = \frac{1}{2} (1-e^{\mp\lambda}) \erfc |v|, \\
\check \psi(\lambda) \equi{t\to\infty} \sqrt\frac{2t}{\pi} \left[\cosh\lambda - 1\right].
\end{align}
We recover the cumulant-generating function of Ref.~\cite{Illien_2013}.

\subsection{Low density}
The opposite limit of low density, $\rho\to 0$, is trickier to define. One should consider it keeping $z = \rho r$ and $\tau = \rho^2 t$ constant. With these scalings, one realizes that the correct limits are
\begin{align}
    \label{eq:ScalingsPhiLowDens}
	\hat \Phi(v, \hat\lambda) &= \lim_{\rho\to 0} \frac{\Phi(v, \lambda = \rho\hat\lambda)}{\rho}, &
	\beta(\hat\lambda) &= \lim_{\rho\to 0} \frac{\pm b_\pm(\lambda =\rho\hat\lambda)}{\rho} = \lim_{\rho\to 0} \frac{A(\lambda =\rho\hat\lambda)}{\rho\hat\lambda}, \\
	\hat F_\mu(v, \hat\lambda) &= \lim_{\rho\to 0} \frac{F_\mu(v, \lambda = \rho\hat\lambda)}{\rho}, &
	\hat G_\mu(v, \hat\lambda) &= \lim_{\rho\to 0} G_\mu(v, \lambda = \rho\hat\lambda). \\
\end{align}

The bulk equation~\eqref{eq:sm_decoup_1b} gives
\begin{equation}
\hat\Phi''(v) + 2(v + \beta)\hat\Phi'(v) + \hat\lambda\left[\hat G_1(v) - \hat G_{-1}(v) \right] = 0.
\end{equation}
This time the equation is not closed. We put forward the following closure relation,
\begin{equation} \label{eq:decoup_difG_low}
\hat G_1(\hat\lambda, v) - \hat G_{-1}(\hat\lambda, v) =
2\frac{d\beta}{d\hat\lambda} \hat \Phi'(v),
\end{equation}
which leads us to
\begin{equation} \label{eq:decoup_1_low_c}
\hat \Phi''(v) + 2(v+\xi) \hat\Phi'(v) = 0,
\end{equation}
with $\xi$ the (rescaled) derivative of the cumulant-generating function with respect to its parameter,
\begin{equation}
\label{eq:xiLowDens}
\xi \equiv \beta + \hat\lambda \frac{d\beta}{d\hat\lambda} = \frac{d}{d\hat\lambda}(\hat\lambda \beta) =  \frac{d\hat A(\hat\lambda, \tau)}{d\hat\lambda}.
\end{equation}

The low density limit of Equations \eqref{eq:sm_decoup_3b}, \eqref{eq:sm_decoup_2b} and \eqref{eq:sm_decoup_4b} is readily taken and the set of equations we need to solve is
\begin{align}
&\hat\Phi''(v) + 2(v + \xi) \hat \Phi'(v) = 0, \\
&\hat\Phi'(0^\nu) + 2\beta\left[1 + \hat\Phi(0^\nu)\right] = 0, \\
\label{eq:sm_decoup_1c}
&\hat\Phi(0^+) - \hat\Phi(0^-) = \hat\lambda, \\
&\hat\Phi(\pm\infty) = 0.
\end{align}
The computation leads to
\begin{equation}
\hat \Phi(v\gtrless 0) = \frac{\pm\beta}{\pi^{-1/2}e^{-\xi^2}\mp \beta\erfc(\pm \xi)}\erfc(\pm(v+\xi))
\end{equation}
Therefore Eq.~\eqref{eq:sm_decoup_1c} yields an implicit equation for $\beta$ and $\xi$,
\begin{equation} \label{eq:sm_low_implicit}
\beta\left(\frac{\erfc(\xi)}{\pi^{-1/2}e^{-\xi^2}-\beta\erfc(\xi)} + \frac{\erfc(-\xi)}{\pi^{-1/2}e^{-\xi^2}+\beta\erfc(-\xi)}\right)
= \hat \lambda. 
\end{equation}
The coefficients $\hat \kappa_1$, $\hat \kappa_2$, \dots{} involved in the cumulants
are defined by
\begin{align}
\label{eq:defCumulLowDens}
\hat\Psi(\hat\lambda, t) &\equiv \sum_{n=1}^\infty \frac{\hat\lambda^n}{n!} \hat\kappa_n \sqrt{2t}, &
\beta &= \sum_{n=0}^\infty \frac{\hat\lambda^n}{(n+1)!} \hat\kappa_{n+1}, &
\xi &= \sum_{n=0}^\infty \frac{\hat\lambda^n}{n!} \hat\kappa_{n+1}.
\end{align}
These expressions can be injected into Eq~\eqref{eq:sm_low_implicit} to obtain the cumulants order by order,
\begin{align} \label{eq:decoup_cum2_low}
	\hat\kappa_2 &= \frac{1}{\sqrt\pi}, &
	\hat\kappa_4 &= \frac{3(4-\pi)}{\pi^{3/2}}, \\
	\hat\kappa_6 &= \frac{15(68-30\pi+3\pi^2)}{\pi^{5/2}}, &
	\hat\kappa_8 &= \frac{21(10912-6840\pi+1320\pi^2-75\pi^3)}{\pi^{7/2}}.
\end{align}
The cumulants are $\kappa_n(t) = \rho^{1-n} \hat\kappa_n \sqrt{2t}$. These are exactly the coefficients known in the literature for interacting point-like particles on a line~\cite{Sadhu_2015, Krapivsky_2014, Hegde_2014}, a model which is equivalent to the low density SEP.
Furthermore, we are able to give the generalized profiles at all orders in $\lambda$,
\begin{align}
	\hat\Phi^{(1)}(v) &= \frac{1}{2}\erfc v,\\ 
	\hat\Phi^{(2)}(v) &= \frac{1}{2}\erfc v - 2\frac{e^{-v^2}}{\pi}, \\
	\hat\Phi^{(3)}(v) &= \frac{3}{\pi^{3/2}}\left[(2v - \sqrt\pi)e^{-v^2} + \sqrt\pi \erfc v \right], \\
	\hat\Phi^{(4)}(v) &= -\frac{1}{2\pi^2}  \left[
	(128 -24\pi +24\sqrt\pi v + 32 v^2)e^{-v^2} + 3\pi (\pi - 8)\erfc v
	\right].
\end{align}

\subsection{Initial step density in the dilute limit}

Our formalism can be applied to the case of an initial step density with $\rho_-$ for $r < 0$ and $\rho_+$ for $r > 0$. At large times, the GDPs take the form
\begin{equation}
    w_r(\lambda, t) - \rho_\nu
    \equi{t\to\infty}  
    \Phi\left(v = \frac{r}{\sqrt{2t}}, \lambda\right)
    \:,
\end{equation}
with $\nu = \sign(r)$. We denote $\rho = (\rho_+ + \rho_-)/2$ the mean density and $\hat{\rho}_\pm = \rho_\pm/\rho$.  In the dilute limit $\rho \to 0$, we obtain the following set of equations
\begin{align}
&\hat\Phi''(v) + 2(v + \xi) \hat \Phi'(v) = 0, \\
&\hat\Phi'(0^\nu) + 2\beta\left[\hat{\rho}_\nu + \hat\Phi(0^\nu)\right] = 0, \\
\label{eq:sm_decoup_1cStep}
&\hat\Phi(0^+) - \hat\Phi(0^-) + \hat{\rho}_+ - \hat{\rho}_- = \hat\lambda, \\
&\hat\Phi(\pm\infty) = 0.
\end{align}
where $\hat{\Phi}$, $\beta$ and $\xi$ are defined by Equations~\eqref{eq:ScalingsPhiLowDens} and~\eqref{eq:xiLowDens}. This leads to
\begin{equation}
\label{eq:ProfStepDens}
\hat \Phi(v\gtrless 0) = \hat{\rho}_\pm \frac{\pm\beta}{\pi^{-1/2}e^{-\xi^2}\mp \beta\erfc(\pm \xi)}\erfc(\pm(v+\xi))
\:.
\end{equation}
Therefore Eq.~\eqref{eq:sm_decoup_1cStep} yields the following implicit equation for $\beta$ and $\xi$,
\begin{equation}
\beta\left(\hat{\rho}_+ \frac{\erfc(\xi)}{\pi^{-1/2}e^{-\xi^2}-\beta\erfc(\xi)} + \hat{\rho}_-\frac{\erfc(-\xi)}{\pi^{-1/2}e^{-\xi^2}+\beta\erfc(-\xi)}\right)
+ \hat{\rho}_+ - \hat{\rho}_-
= \hat \lambda
\:. 
\end{equation}
We can deduce the coefficients $\hat \kappa_1$, $\hat \kappa_2$, \dots{} involved in the cumulants by using the expansions of $\xi$ and $\beta$~\eqref{eq:defCumulLowDens}. We obtain that $\hat{\kappa}_1$ is solution of
\begin{equation}
\frac{\hat{\rho}_+}{1-\sqrt{\pi} \ed^{\hat{\kappa}_1^2} \erfc(\hat{\kappa}_1)}
= \frac{\hat{\rho}_-}{1+\sqrt{\pi} \ed^{\hat{\kappa}_1^2} \erfc(-\hat{\kappa}_1)}
\:. 
\end{equation}
The higher cumulants can be expressed in terms of $\hat{\kappa}_1$. For instance,
\begin{equation}
\hat{\kappa}_2
= \frac{2(1-\sqrt{\pi} \ed^{\hat{\kappa}_1^2} \erfc(\hat{\kappa}_1))(1+\sqrt{\pi} \ed^{\hat{\kappa}_1^2} \erfc(-\hat{\kappa}_1))}{\hat{\rho}_+(4\hat{\kappa}_1+\sqrt{\pi} (1+4\hat{\kappa}_1^2) \ed^{\hat{\kappa}_1^2} \erfc(-\hat{\kappa}_1)) - \hat{\rho}_-(4\hat{\kappa}_1-\sqrt{\pi} (1+4\hat{\kappa}_1^2) \ed^{\hat{\kappa}_1^2} \erfc(\hat{\kappa}_1))}
\:. 
\end{equation}
The expressions of $\hat{\kappa}_1$ and $\hat{\kappa}_2$ are identical to the exact ones obtained previously~\cite{Landim:1998a,Imamura:2017} in the dilute limit. In addition to the cumulants, we obtain the expressions of the density profiles at any order, by expanding~\eqref{eq:ProfStepDens} in powers of $\hat{\lambda}$. For instance, for $v>0$,
\begin{align}
	\hat\Phi^{(1)}(v) &= \frac{\hat{\kappa}_2 \hat{\rho}_+ \sqrt{\pi} \ed^{\hat{\kappa}_1^2}}{2(1-\sqrt{\pi} \ed^{\hat{\kappa}_1^2} \hat{\kappa}_1 \erfc(\hat{\kappa}_1))^2} \erfc(v+\hat{\kappa}_1)
	- \frac{2 \hat{\kappa}_1 \hat{\kappa}_2 \hat{\rho}_+}{1-\sqrt{\pi} \ed^{\hat{\kappa}_1^2} \hat{\kappa}_1 \erfc(\hat{\kappa}_1)} \ed^{-v(v+\hat{\kappa}_1)}
	\:.
\end{align}

\subsection{Solution at all times in the high density limit}

In the high density limit $\rho\to 1$, the cumulant-generating function
is expected to scale as $(1-\rho)$. We write
\begin{align}
\psi(\lambda, t) &\equi{\rho\to 1} (1-\rho) \check\psi(\lambda, t), &
B_\mu(\lambda, t)  &\equi{\rho\to 1} (1-\rho) \check B_\mu(\lambda, t)  = (1-\rho)\frac{\partial_t \check \psi(\lambda, t) }{e^{\mu\lambda} - 1},
\end{align}
with $\check\psi$ and $\check B_\mu$ independent of the density $\rho$.

The fluctuations of occupation $\delta\eta_r = \eta_r - \langle \eta_r\rangle$ also scale as $(1-\rho)$. Thus, the generalized profiles $w_r$ scale as $(1-\rho)$ while the correlations $f_{\mu, r}$ (between $\eta_{X+\mu}$ and $\eta_{X+r}$) scale as $(1-\rho)^2$.
\begin{align}
w_r &\equi{\rho\to 1} \rho + (1-\rho) \check w_r = 1 + (1-\rho)(\check w_r - 1), \\
f_{\mu, r} &= \mathcal{O}[(1-\rho)^2].
\end{align}

When all the scalings are written, the microscopic equations~\eqref{eq:sm_evolPsi}-\eqref{eq:decoup_2} become a closed system independent of $\rho$,
\begin{align}
\label{eq:decoup_1_high}
\partial_t \check w_r &= \frac{1}{2} \Delta \check w_r \\
\label{eq:decoup_2_high}
\partial_t \check w_\mu &= \frac{1}{2} \nabla_\mu \check w_\mu + \check B_\nu(t) \\
\label{eq:decoup_3_high}
\lim_{r\to\pm\infty} \check w_r &= 0 \\
\label{eq:decoup_4_high}
  \partial_t \check\psi &= \frac{1}{2} \left[
                          (e^\lambda - 1)(1 - \check w_1) + (e^{-\lambda} - 1)(1 - \check w_{-1}) \right]
\end{align}

We define the Laplace transform
\begin{equation}
\tilde w_r(u) = \int_0^\infty  e^{-ut} \check w_r(t).
\end{equation}
The bulk and boundary equations become,
\begin{align}
\label{eq:decoup_1_high_2}
\frac{1}{2} \left[\tilde w_{r+1}(u) +\tilde w_{r-1} (u)\right] - (1+u) \tilde w_r(u) = 0 \\
\label{eq:decoup_2_high_2}
\frac{1}{2} \tilde w_{2\nu}(u) - \left(\frac{1}{2}+u\right) \tilde w_\nu(s) + \nu \tilde B_\nu(u) = 0.
\end{align}
The equation $\alpha^2 - 2(1+u) \alpha + 1 = 0$ has two solutions, but only one satisfies the condition $\alpha^r\underset{r\to\infty}{\to} 0$ imposed by Eq.~\eqref{eq:decoup_3_high}. The solution of Eq.~\eqref{eq:decoup_1_high_2} is
\begin{gather}
\tilde w_r(u) = \gamma_\mu(u) \alpha^{|r|}, \\
\alpha = 1+u-\sqrt{(1+u)^2-1},
\end{gather}
where $\mu$ is the sign of $r$.
Injecting this expression into the boundary equation~\eqref{eq:decoup_2_high_2}, we obtain (recall that $\alpha^2 - 2(1+u)\alpha + 1 = 0$)
\begin{equation}
\gamma_\mu(u) = \frac{2\tilde B_\mu(u)}{(1+2u)\alpha - \alpha^2} = \frac{2 \tilde B_\mu(u)}{1-\alpha}
= \frac{2}{1-\alpha} \frac{(\partial_t\tilde\psi)(u)}{e^{\mu\lambda} - 1}.
\end{equation}
We finally use the velocity equation~\eqref{eq:decoup_4_high} and obtain
\begin{gather}
(\partial_t\tilde\psi)(u) = \frac{1}{2u}\left(e^{\lambda} + e^{-\lambda} - 2\right) - \frac{2\alpha}{1-\alpha}(\partial_t\tilde\psi)(u), \\
(\partial_t\tilde\psi)(u) = \frac{1}{u} \frac{1-\alpha}{1+\alpha} \left[\cosh\lambda -1\right]
= \frac{1}{\sqrt{u(2+u)}} \left[\cosh\lambda -1\right].
\end{gather}
This expression can be inverted into
\begin{equation}
\partial_t\check\psi(t) = e^{-t} I_0(t) \left[\cosh\lambda -1 \right].
\end{equation}
The large time limit is given by
\begin{equation}
\partial_t\check\psi(t) \equi{t\to\infty} \frac{1}{\sqrt{2\pi t}} \left[\cosh\lambda -1 \right].
\end{equation}

We also obtain the full solution for the generalized profiles $\tilde w_r$,
\begin{equation} \label{eq:decoup_solWr_high}
\tilde w_r(u) = \frac{1}{u} \frac{1}{1+\alpha}\left[1 - e^{-\mu\lambda} \right] \alpha^{|r|}.
\end{equation}
The small $u$ behavior at constant $r\sqrt{u}$ gives the large time behavior at constant $r/\sqrt{t}$,
\begin{align}
\tilde w_r(u) &\equi{u\to 0}  \left[1 - e^{-\mu\lambda} \right] \frac{e^{-|r|\sqrt{2u}}}{2u} \\ \label{eq:decoup_solWr_high_scale}
\check w_r(t) &\equi{t\to\infty} \frac{1}{2} \left[1 - e^{-\mu\lambda} \right] \erfc\left(\frac{|r|}{\sqrt{2t}}\right).
\end{align}

\section{Generic single-file systems}
\subsection{Description of single-file systems in terms of two quantities}
Two descriptions of single-file systems at large distance and large time have been put forward. They both involve two quantities.

The first description comes from fluctuating hydrodynamics~\cite{Spohn_1983}. The system considered is a lattice model. It is described at large distance and large time by a fluctuating density field $\rho(x, t)$ that is shown to obey the following equation,
\begin{equation}
\partial_t \rho(x, t) = \partial_x\left[D(\rho(x, t)) \partial_x \rho(x, t) + \sqrt{\sigma(\rho(x, t))}\eta(x, t)\right].
\end{equation}
The quantities $D(\rho)$ and $\sigma(\rho)$ were first defined from the microscopic details of a lattice gas~\cite{Spohn_1983}. It is nevertheless more intuitive to consider a system of size $L$ between two reservoirs at densities $\rho_a$ and $\rho_b$~\cite{Derrida_2007}.
The number of particles transferred from left to right at time $t$ is denoted $Q_t$ and is shown to satisfy
\begin{align}
\lim_{t\to\infty} \frac{\left\langle Q_t\right\rangle}{t}
&= \frac{D(\rho)}{L}(\rho_a - \rho_b) \text{\qquad if } (\rho_a - \rho_b) \text{ is small}, &
\lim_{t\to\infty} \frac{\left\langle Q_t\right\rangle}{t}
&= \frac{\sigma(\rho)}{L} \text{\qquad if } \rho_a=\rho_b = \rho.
\end{align}
This can be used as a definition of $D(\rho)$ and $\sigma(\rho)$. Using macroscopic fluctuation theory (MFT), it has been shown~\cite{Krapivsky_2015b} that the variance $\kappa_2$ of the displacement of a tagged particle in the system satisfies
\begin{equation} \label{eq:sm_msd_mft}
\kappa_2 \equi{t\to\infty} \frac{\sigma(\rho)}{\rho^2} \sqrt\frac{t}{\pi D(\rho)}.
\end{equation}

The second description has been developed by Kollmann~\cite{Kollmann_2003}. The system consists of identical Brownian particles with pairwise interactions. Denoting the fluctuating density field $\rho(x, t)$, one defines the dynamical structure factor $S(q, t)$ as
\begin{align}
S(q, t) &= \frac{1}{N} \left\langle\delta \rho(q, t)\delta \rho(q, 0) \right\rangle, &
\delta \rho(q, t) &= \int dx\, e^{iqx} \left[\rho(x, t) - \bar\rho\right],
\end{align}
where $N$ is the number of particles and $\bar\rho$ the average density. The structure factor decays exponentially with time: $S(q, t) \sim S(q, 0) e^{-D(q) t}$.
The behavior of the system is shown to be dominated by the large wavelengths ($q \to 0$). The two important quantities (with their dependence on $\bar\rho$ written explicitly) are
\begin{align}
D(\bar\rho) &= \lim_{q\to 0} D(q), &
S(\bar\rho) &= S(q=0, t=0). 
\end{align}
Kollmann shows that the mean square displacement of a tagged particle in the system satisfies
\begin{equation} \label{eq:sm_msd_kol}
\kappa_2 \equi{t\to\infty} \frac{2S(\rho)}{\rho} \sqrt\frac{D(\rho) t}{\pi}.
\end{equation}

Ref.~\cite{Krapivsky_2015b} provides the link between the two approaches (Eq.~(95)). $D(\rho)$ is the same quantity and $\sigma(\rho)$ and $S(\rho)$ are linked by
$\sigma(\rho) = 2\rho D(\rho)S(\rho)$. Eqs.~\eqref{eq:sm_msd_mft} and \eqref{eq:sm_msd_kol} are thus identical.

In the main text, we chose the description of Kollmann in terms of $D(\rho)$ and $S(\rho)$. We now list the values of these two quantities for the systems that we consider.
\[
\renewcommand{\arraystretch}{2}
\begin{array}{l*2{|>{\displaystyle}c}}
\text{Model} & D(\rho)  & S(\rho)\\ \hline
\text{Symmetric exclusion process~\cite{Krapivsky_2015b}} & D_0 & 1-\rho  \\
\text{Point-like hard core particles~\cite{Krapivsky_2015b}} & D_0 & 1 \\
\text{\parbox{6cm}{Pairwise interacting particles\newline without hydrodynamic interactions~\cite{Kollmann_2003,Lin_2005}}} & \frac{D_0}{S(\rho)} & S(\rho) \\
\text{Hard rod gas~\cite{Lin_2005}} & \frac{D_0}{(1-a\rho)^2} & (1-a\rho)^2\\
\text{Random average process~\cite{Kundu_2016}} & \frac{\mu_1}{2\rho^2} & \frac{\mu_2}{\mu_1 - \mu_2}
\end{array}
\]
$D_0$ is the diffusion coefficient of an individual particle, $a$ is the length of the hard rods, and $\mu_k$ are the moments of the probability law of the jumps in the RAP~\cite{Kundu_2016}. In the case of pairwise interacting particles, the structure factor $S(\rho)$ can either be determined directly from the positions, or indirectly from the pair correlations $g(r)$ via the compressibility relation~\cite{Hansen},
\begin{align}
S(\rho) &= \lim_{q\to 0} \frac{1}{N}\left\langle \sum_{i, j} e^{iq(X_i-X_j)}\right\rangle 
= 1 + \int_{-\infty}^\infty dr \left[g(r) - 1\right]. \label{eq:compress_eq}
\end{align}

\subsection{Extension of our approach}
In light of the description in terms of the two quantities $D(\rho)$ and $S(\rho)$, we extend Eqs.~\eqref{eq:sm_decoup_1d}-\eqref{eq:sm_decoup_4d} to generic single-file systems.
\begin{align}
D(\rho) \Phi_1''(v) + v \Phi_1'(v) = 0, \\
D(\rho)\Phi_1'(0^\pm) + \frac{1}{2}\rho\tilde\kappa_2 = 0, \\
\Phi_1(0^+) - \Phi_1(0^-)= S(\rho), \\
\Phi_1(v) \xrightarrow[v\pm\infty]{} 0,
\end{align}
with $\rho$ the average density of the system.
We recall that for the SEP, we had $D(\rho) = 1/2$ and $S(\rho) = 1-\rho$.

The solution is readily obtained:
\begin{align}
\Phi_1(v\gtrless 0) &= \pm\frac{S(\rho)}{2} \erfc\left(\frac{|v|}{\sqrt{2D(\rho)}}\right), \label{eq:Phi1_general} \\
\tilde\kappa_2 &= \frac{S(\rho)}{\rho} \sqrt\frac{2 D(\rho)}{\pi}.
\end{align}
We stress that this solution is exact, as confirmed by the alternative derivation provided in Section \ref{sec:MFT}. 

If we call $\eta(x, t)$ the density field at $x$ at time $t$ and $X_t$ the position of the tracer at time $t$, this means
\begin{align}
\left\langle\eta(X_t + x, t) X_t\right\rangle &\equi{t\to\infty} \sign(x) \frac{S(\rho)}{2} \erfc\left(\frac{|x|}{\sqrt{4D(\rho) t}}\right) \\
\left\langle X_t^2\right\rangle &\equi{t\to\infty} \frac{S(\rho)}{\rho} \sqrt\frac{4D(\rho) t}{\pi}.
\end{align}
The result for the variance is the one given in Refs~\cite{Kollmann_2003,Krapivsky_2015b}.

\subsection{MFT}
\label{sec:MFT}

In the formalism of Macroscopic Fluctuation Theory (MFT), the main object is the density $\rho(x,\tau)$ which is the continuous equivalent of the occupation $\eta_i(t)$ at time $t = \tau T$, $\tau \in [0,1]$ with $x = i/\sqrt{T}$. The probability to start from a density $\rho_0$ at $t=0$ and end up with a density $\rho(x,1)$ at $t=T$ is given by~\cite{Derrida_2009}:
\begin{equation}
  \mathbb{P}(\rho_0(x) \longrightarrow \rho(x,1))
  = \int \Df [\rho(x,\tau)] \Df[H(x,\tau)] \: \ed^{-\sqrt{T} \: S[\rho,H]}
  \:,
\end{equation}
where the action $S$ reads
\begin{equation}
  \label{eq:ActS}
  S[\rho,H] =
  \int \dd x \int_0^1 \dd \tau
  \left(
    H \partial_\tau \rho + \frac{1}{2} \partial_x \rho \partial_x H
    - \frac{\rho(1-\rho)}{2} (\partial_x H)^2
  \right)
  \:.
\end{equation}
The distribution of the initial condition $\rho_0$ takes the form
\begin{equation}
  \mathbb{P}[\rho_0] \simeq \ed^{- \sqrt{T} \: F[\rho_0]}
  \:,
\end{equation}
where
\begin{equation}
  \label{eq:ActF}
  F[\rho(x,0)] = \int \dd x \int_\rho^{\rho(x,0)} \dd z \: \frac{\rho(x,0)-z}{z(1-z)}
  \:.
\end{equation}

The cumulant generating function for the position of the tracer can be written as~\cite{Krapivsky_2015}
\begin{equation}
\label{eq:DefCumulGenFctMFT}
  \left\langle \ed^{\lambda X_T} \right\rangle \simeq \int \Df \rho_0
  \int \Df [\rho(x,\tau)] \Df[H(x,\tau)] \: \ed^{-\sqrt{T} \: (S[\rho,H] + F[\rho_0] - \lambda Y[\rho])}
  \:,
\end{equation}
where $Y[\rho] = X_T/\sqrt{T}$ is the position of the tracer. It is deduced from $\rho(x,\tau)$ from the conservation of the number of particles to the right of the tracer:
\begin{equation}
  \label{eq:FoncXt}
  \int_0^{Y[\rho]} \rho(x,1) \dd x
  = \int_0^{\infty} \left( \rho(x,1) - \rho(x,0) \right) \dd x
  \:.
\end{equation}
For large $T$, the integral in~\eqref{eq:DefCumulGenFctMFT} is dominated by the minimum of $S+F-\lambda Y$, taken as a function of $(\rho,H)$. We denote this minimum $(q,p)$. These functions satisfy the equations~\cite{Krapivsky_2015}
\begin{align}
  \label{eq:MFT_q}
  \partial_\tau q &= \partial_x[D(q) \partial_x q] - \partial_x[\sigma(q)\partial_x p]
  \:,
  \\
  \label{eq:MFT_p}
  \partial_\tau p &= - D(q) \partial_x^2 p - \frac{1}{2}  \sigma'(q) (\partial_x p)^2 
  \:,
\end{align}
with the terminal condition for $p$
\begin{equation}
  \label{eq:MFT_limitP}
  p(x,\tau=1) = B \Theta(x-Y)
  \:,
  \quad B = \frac{\lambda}{q(Y,1)}
  \:,
\end{equation}
and the initial condition for $q$, expressed in terms of $p(x,0)$:
\begin{equation}
  \label{eq:MFT_limitQ}
  p(x,0) = B \Theta(x) + \int_{\rho}^{q(x,0)} \dd r \frac{2 D(r)}{\sigma(r)}
  \:.
\end{equation}
This approach has been used to compute the first four cumulants of the position of the tracer~\cite{Krapivsky_2015}.

We can make a connection between this approach and the generalized profiles, since the latter can be expressed as
\begin{equation}
  \label{eq:FctIntegW}
  w_r(\lambda,T) = \frac{\left\langle \eta_{X_T+r}(T) \ed^{\lambda X_T} \right\rangle}{\left\langle \ed^{\lambda X_T} \right\rangle}
  \simeq \frac{
    \displaystyle
    \int \Df \rho_0
    \int \Df [\rho(x,\tau)] \Df[H(x,\tau)] \: \rho(r/\sqrt{T} + Y[\rho],1) \: \ed^{-\sqrt{T} \: (S[\rho,H] +
    F[\rho_0] - \lambda Y_T[\rho])}
  }
  {
    \displaystyle
    \int \Df \rho_0
    \int \Df [\rho(x,\tau)] \Df[H(x,\tau)] \: \ed^{-\sqrt{T} \: (S[\rho,H] +
      F[\rho_0] - \lambda Y_T[\rho])}
  }
  \:.
\end{equation}
The two integrals can be evaluated via a saddle point method. The saddle point is the same for the numerator and the denominator and is given by $(q,p)$ solution of~(\ref{eq:MFT_q},\ref{eq:MFT_p}). Therefore, for large $T$,
\begin{equation}
  \label{eq:EquivMFTprof0}
  w_r(\lambda,T)
  \simeq q\left(\frac{r}{\sqrt{T}} + Y[q],\tau=1 \right)
  \:.
\end{equation}
In our formalism we have $v = r/\sqrt{2T}$, thus,
\begin{equation}
  \label{eq:EquivMFTprof}
    \Phi(v) = q(v \sqrt{2}+Y[q], \tau=1) - \rho
    \:,
\end{equation}
which relates our generalized profiles to the MFT solution.

The MFT equations~(\ref{eq:MFT_q},\ref{eq:MFT_p}) can be solved perturbatively at first order in $\lambda$. Let us denote
\begin{align}
  \label{eq:expQ}
  q(x,\tau) &= \rho + \lambda \: q_1(x,\tau)
              + \cdots \:,
  \\
  \label{eq:expP}
  p(x,\tau) &= \lambda \: p_1(x,\tau)
              + \cdots  \:.
\end{align}
We also write the expansion of $Y[q]$:
\begin{equation}
  \label{eq:ExpYB}
  Y = Y_0 + \lambda \: Y_1 + \cdots \:.
\end{equation}
The coefficients can be determined via the condition~(\ref{eq:FoncXt}) which yields
\begin{align}
  \label{eq:Yorder0}
  \rho Y_0
  &= 0
    \:,
  \\
  \label{eq:Yorder1}
  \rho Y_1
  &= \int_0^\infty \left( q_1(x,1) - q_1(x,0) \right) \dd x
    \:.
\end{align}
The boundary conditions~(\ref{eq:MFT_limitP},\ref{eq:MFT_limitQ}) give a series of conditions for the $p_i$'s and $q_i$'s. At first order, we get
\begin{align}
    \label{eq:TermCdtP1}
    p_1(x,1) &= \frac{1}{\rho} \Theta(x) \:,
    \\
   \label{eq:InitCdtQ1}
    q_1(x,0) &= \frac{\sigma(\rho)}{2 D(\rho)}(p_1(x,0) - \rho^{-1} \Theta(x)) \:,
\end{align}
and the MFT equations~(\ref{eq:MFT_q},\ref{eq:MFT_p}) become
\begin{align}
  \label{eq:MFT_q1}
  \partial_\tau q_1 &= D(\rho) \partial_x^2 q_1 -
                 \sigma(\rho) \partial_x^2 p_1
                 \:.
  \\
  \label{eq:MFT_p1}
  \partial_\tau p_1 &= - D(\rho) \partial_x^2 p_1 \:,
\end{align}
We first solve the equation for $p_1$ and then use the result to solve the equation for $q_1$. This gives, at $\tau=1$:
\begin{equation}
  q_1(x,\tau=1) =  \frac{\sigma(\rho)}{4 \rho D(\rho)} \sign(x)
  \erfc \left(
    \frac{|x|}{\sqrt{4 D(\rho)}}
  \right)
  \:.
\end{equation}
From~(\ref{eq:EquivMFTprof}), we deduce
\begin{equation}
    \Phi_1(v) =  \frac{\sigma(\rho)}{4 \rho D(\rho)} \sign(v)
    \erfc \left(
      \frac{|v|}{\sqrt{2D(\rho)}}
    \right)
    \:,
\end{equation}
which is exactly \eqref{eq:Phi1_general}.

\section{Numerical simulations}

\subsection{Symmetric exclusion principle} Simulations of the SEP are performed on a periodic ring of size $N=1000$. The average density is set to $\rho$ and $M=\rho N$ particles are initially placed uniformly at random on the ring. The successive jumps of the particles are implemented as follow: one chooses a particle uniformly at random and one of the two possible directions (left and right) with equal probabilities. If the chosen particle has no neighbor in the chosen direction, the jump is performed, else it is rejected. In both cases, the time of the simulation is incremented by a random number drawn according to an exponential distribution of rate $N$.

We keep track of one particle (the tracer) and compute its moments and the generalized profiles at the times that we want. The average is taken over $10^8$ repetitions of the simulation.

\subsection{Point-like hard-core diffusive particles}
We consider particle that diffuse on a line, with hard-core exclusions. One notes that the dependence in the density is trivial since at density $\rho$ the space $x$ and the time $\tau$ can be rescaled as
$x \gets \rho x$ and $\tau \gets \rho^2\tau$. We thus consider only $\rho = 1$. Initially ($\tau=0$), $M=20001$ particles are placed uniformly at random on the interval $[0, M]$. The tracer is particle number $(M+1)/2$. We consider that the particles diffuse independently of one another, then we implement the hard core interactions by restoring the order. Practically, between time $0$ and $\tau$, particle $k$ moves by $\Delta x_i$ drawn according to the Gaussian probability law
\begin{equation}
P(\Delta x_k, \tau) = \frac{1}{\sqrt{2\pi\tau}} e^{-\frac{(\Delta x_k)^2}{2\tau}}.
\end{equation}
The tracer is still particle number $(M+1)/2$ from the left. Its displacement and the density field in its reference frame are easily computed. To compute the observables, the average is performed over $4\cdot 10^8$ repetitions.

\subsection{Hard-rod gas}
We consider a gas of diffusive hard rods of size $a<1$ at density $\rho=1$. The position of rod $k$ at time $t$ is denoted $y_k(t)$.
We may substract the rod sizes from the positions and define $x_k(t)=y_k(t) - ka$. One realizes that the set $\{x_k(t)\}$ correspond to point-like hard-core diffusive particles.
Using the mapping $y_k(t) = x_k(t) + k a$ at both the initial and final times, one can compute the observables of the hard-rod gas from simulations of the point-like particles described above.

The low-density limit of the model correspond exactly to the point-like hard-core diffusive particles. In this case, the profiles at order $2$ and $3$ are given in Fig.~3 of the article.

\subsection{Random-average process}
We consider the random-average process (RAP) defined in particular in Ref.~\cite{Kundu_2016}. Particles are placed on the infinite one-dimensional line. They are all embedded with exponential clocks of characteristic time $1$. When its clock ticks a particle jumps choose a direction, left or right, with equal probability. It then jumps in this direction at a distance which corresponds to a fraction $\eta$ of the distance to its nearest neighbor. $\eta$ is a random variable following a probability law on $[0, 1]$. In our simulations, we choose the uniform probability law.

By construction of the RAP, if the density of the particles is denoted $\rho$ and if $x$ and $t$ are respectively the spatial and temporal coordinates, the observables depend only on the two rescaled coordinates $z = \rho x$ and $\tau = \rho t$. For this reason, we only consider the RAP at density $\rho = 1$.

In our simulations, we consider $N=500$ particles on a periodic line (of length $L = 500$). The steady state of the RAP is non-trivial~\cite{Kundu_2016} and can hardly be implemented as an initial condition. We thus first let the system evolve for a time $t_\mathrm{ini} = 2\cdot 10^4$ before starting to record the observables. These observables are then averaged over $2 \cdot 10^6$ simulations.

Note that the low-density limit of the RAP is peculiar since $S(\rho)\sim \rho^{-1}$. It does not correspond to the ideal gas ($S(\rho) = 1$). Therefore, the profiles at order $2$ cannot be checked against our low-density prediction.

\subsection{Point-like particles interacting by a pairwise potential}
We consider $N$ particles ($N=500$ for WCA potential, $N=200$ for dipole-dipole potential) on a ring of length $L=N/\rho$ where $\rho$ is the density. The particles diffuse with a diffusion coefficient $D_0 = 1$. In addition, they interact by a pair potential $V(r)$. Two kinds of interactions are considered: a short-range WCA potential $V_\mathrm{WCA}(r)$ and a long-range dipole-dipole potential $V_\mathrm{dip}(r)$,
\begin{align}
V_\mathrm{WCA}(r) &=
\begin{cases} 
4A_\mathrm{WCA}\left(\frac{1}{r^{12}} - \frac{1}{r^6}\right) & r < 2^{1/6} \\
0 & r > 2^{1/6}
\end{cases}, &
V_\mathrm{dip}(r) &= \frac{A_\mathrm{dip}}{r^3},
\end{align}
with $A_\mathrm{WCA} = A_\mathrm{dip} = 1$. In the case of the dipole-dipole interaction, we consider that the ring is a circle of radius $L/(2\pi)$ embedded in 2d space: the distance $r$ between two particle is the distance between the points of the circles so that the 1d force is the tangential component of the 2d force.

The time-step for the Brownian dynamics
is set to $\Delta t =2\cdot 10^{-4}$. Starting from random initial positions, we let the system equilibrate during a time $t_0 = 10^4$ before recording the observables. At each iteration, we check that the particles are ordered and if they are not, we restart the simulation ($< 1\%$ of simulations in the worst case). To regularize the diverging potentials at small distance, we ensure a maximum displacement of a particle during a time iteration ($0.1$ units ; only a frequency $\sim 10^{-7}$ of the moves need this regularization).
 We average the results over 50000 simulations (WCA potential) or 2500 simulations (dipole-dipole potential).
 
\begin{figure}
    \includegraphics[scale=1]{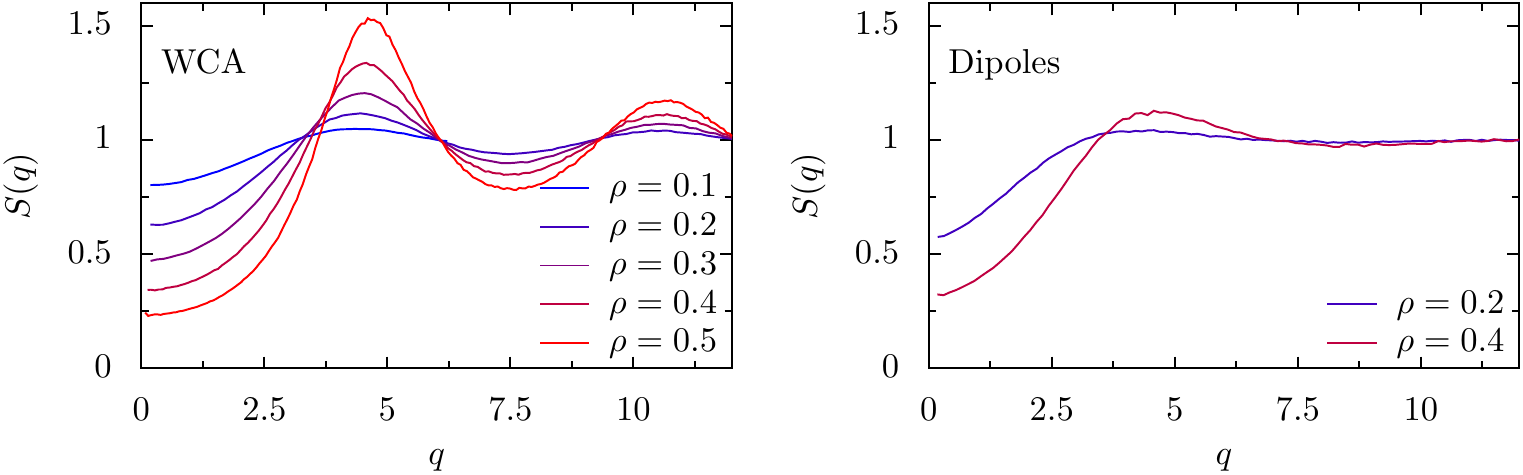} \\[0.3cm]
    \begin{tabular}{c|c|c|c|c|c}
		$\rho$  & 0.1 & 0.2 & 0.3 & 0.4 & 0.5 \\ \hline
		$S(\rho) = S(q=0)$ & 0.82 & 0.63 & 0.48 & 0.35 & 0.23
	\end{tabular}
	\qquad\qquad\qquad
	\begin{tabular}{c|c|c}
		$\rho$  & 0.2 & 0.4 \\ \hline
		$S(\rho)  = S(q=0)$ & 0.58 & 0.32
	\end{tabular}
	\phantom{AAAAAAAAAA}
	\caption{Structure factor $S(q)$ and measured value of $S(\rho) = S(q=0)$. Left: WCA potential. Right: dipole-dipole potential.}
	\label{fig:sm_sq}
\end{figure}

The structure factor $S(q)$ used in Fig. 4 of the article is computed by the two methods of Eq.~\eqref{eq:compress_eq} (we check that they are consistent). The graphics and the table of values are given in Fig.~\ref{fig:sm_sq}.

